\documentclass[aps,english,10pt,a4,superscriptaddress,onecolumn,nofootinbib]{revtex4}  
\usepackage{amsmath}
\usepackage{amsthm}
\usepackage{amssymb}
\usepackage{amsfonts}
\usepackage{bbm}
\usepackage{epsfig}
\usepackage{times}
\usepackage{babel}
\usepackage{color}
\usepackage{hyperref}
\usepackage{framed}
\usepackage{changes}
\usepackage{soul}

\def\tr{{\rm tr}}

\def\R{{\mathbb{R}}}

\def\be{\begin{equation}}
\def\ee{\end{equation}}
\def\ba{\begin{eqnarray}}
\def\ea{\end{eqnarray}}
\def\R{\mathbb{R}}

\newcommand\nn{\nonumber}

\newtheorem{theorem}{Theorem}
\newtheorem{definition}[theorem]{Definition}

\begin{document}
\title{Is the local linearity of space-time inherited from the linearity of probabilities?}
\author{Markus P.\ M\"uller\footnotemark[2]}
\affiliation{Departments of Applied Mathematics and Philosophy, University of Western Ontario, London, ON N6A 5BY, Canada}
\affiliation{Perimeter Institute for Theoretical Physics, Waterloo, ON N2L 2Y5, Canada}
\author{Sylvain Carrozza\footnotemark[2]}
\affiliation{Universit\'e de Bordeaux, LaBRI, UMR 5800, 33400 Talence, France}
\affiliation{Perimeter Institute for Theoretical Physics, Waterloo, ON N2L 2Y5, Canada}
\author{Philipp A.\ H\"ohn}
\affiliation{Vienna Center for Quantum Science and Technology, and Institute for Quantum Optics and Quantum Information, Austrian Academy of Sciences, Boltzmanngasse 3, 1090 Vienna, Austria}

\begin{abstract}
The appearance of linear spaces, describing physical quantities by vectors and tensors, is ubiquitous in all of physics, from classical mechanics to the modern notion of local Lorentz invariance. However, as natural as this seems to the physicist, most computer scientists would argue that something like a ``local linear tangent space'' is not very typical and in fact a quite surprising property of any conceivable world or algorithm. In this paper, we take the perspective of the computer scientist seriously, and ask whether there could be any inherently information-theoretic reason to expect this notion of linearity to appear in physics. We give a series of simple arguments, spanning quantum information theory, group representation theory, and renormalization in quantum gravity, that supports a surprising thesis: namely, that the local linearity of space-time might ultimately be a consequence of the linearity of probabilities. While our arguments involve a fair amount of speculation, they have the virtue of being independent of any detailed assumptions on quantum gravity, and they are in harmony with several independent recent ideas on emergent space-time in high-energy physics.
\end{abstract}

\date{January 11, 2017}

\maketitle

\section{Introduction}
\label{SecIntroduction}
The notion of vectors and linearity is ubiquitous in physics, and usually considered a fundamental and natural aspect of our world. In classical mechanics, the consensus seems to be that the vectorial description of quantities like position or velocity is a completely unsurprising triviality\footnote{It clearly \textit{is} a triviality if we posit Hamiltonian mechanics on the usual phase space; the observation is more of a pedagogical kind: when classical mechanics is taught, not much effort is put into the justification of a vector space structure, whereas the notion of a symplectic structure, for example, is usually considered counterintuitive and introduced with a lot more pedagogical effort. Even if it is not directly reflected by the mathematics, the notion of naturalness, and the intuition as to which mathematical structures are natural and which demand explanation, has a large impact on theory building in theoretical physics.}. In modern physics, this paradigm has been replaced, and reduced to, a notion of \textit{local Lorentz invariance}, which is not only used as a principle in the construction of pillars like quantum field theory, but has also passed stringent experimental tests that rule out moderate violations as predicted by some versions of quantum gravity~\cite{Mattingly}.

However, given the prominent role of vectors, linearity and symmetries in physics, one may as well take one step back and start to wonder: is it really \textit{that} natural to expect these notions to show up in physics? Thinking back at school times, one of the authors of this paper remembers having been really surprised by his first encounter with vectors in classical mechanics. Suppose one particle moves with velocity $u=(u_1,u_2,u_3)$, and another particles moves with velocity $v=(v_1,v_2,v_3)$ relative to the first particle. Then the second particle's total velocity is $u+v$. \textit{Really? We only add up the numbers, one after the other, na\"ively?} A strong sense of skepticism was felt against this simple recipe -- skepticism that got even worse with the introduction of rotation matrices, and that was built on the experience with other possible worlds: the worlds of computer games.

Think of one favorite possible world of the 1980's, namely the Pacman universe\footnote{To have a more formal version of the argument, we should instead consider (agents within) algorithms that do not rely on external input by a player, such as Conway's ``Game of Life''. However, we here stick to Pacman for illustrative reasons.}. In this popular arcade game, a cake-shaped yellow protagonist is supposed to eat all yellow dots in a labyrinth-like 2D square grid, while avoiding the encounter of dangerous coloured ghost-like creatures. This seems to be a perfectly conceivable world, like ours (only way more boring) -- but it does not contain anything like vectors. Yes, we can count pixels and specify the position of a ghost in terms of two integers, but there is no notion of $\mathbb{R}^n$, vector addition, or Euclidean structure. One may even argue that the very presence of a rudimentary vector structure (two integers) is simply an artifact of the fact that we build our computer monitors in a way to depict familiar (vectorial) structure of our own world. If we consider all possible \textit{algorithms} in the sense of theoretical computer science, then the presence of a vector space structure over the real numbers seems to be an extremely rare phenomenon which has to be built in by hand.

Or so goes the argumentation of a teenage computer nerd in the 1990's at his first encounter of classical mechanics. Ironically, modern physics supports the nerd's point of view due to several recent developments. First, Jacobson's derivation of the Einstein equation as an equation of state~\cite{Jacobson} or from entanglement equilibrium conditions \cite{ted}, and the subsequent developments regarding holographic duality \cite{bianchi,swingle2,raamsdonk}, have led to the idea that the dynamics of spacetime itself may emerge from an underlying level involving (quantum) information theory. Second, research in quantum gravity suggests in many ways that our notion of smooth spacetime manifold is most likely not fundamental, but an emergent approximation to an underlying quantum theory of gravity. The geometric architecture of general relativity, including the local (linear) tangent space structure of space-time, may therefore only be the low-energy manifestation of the collective behaviour of quantum and pre-geometric degrees of freedom.
Taking information-theoretic notions among the primary ingredients, and spacetime as emergent, it is thus a non-trivial open problem to understand the properties of the latter in terms of the former.

This paper starts with a simple, yet powerful observation. Consider a possible world like the Pacman universe. Even if there is no notion of vectors in the usual sense, there \textit{is} a notion of linearity -- namely, linearity in probabilities. Concretely, as soon as we have an agent in that universe (say, Pacman) who has incomplete knowledge of his world (is there a ghost around the corner?), that agent will use probability theory to organize his knowledge. For example, if the agent assigns probabilities $p(B_n)$ to an exhaustive list of mutually exclusive events $B_1,B_2,\ldots$, then the probability he should assign to any other event $A$ is a linear functional of the vector $(p(B_1),p(B_2),\ldots)$, namely
\[
   p(A)=\sum_n p(A|B_n)p(B_n).
\]
This observation suggests a bold speculation: given the fact that it is a generic property of possible worlds that agents use probability theory to organize their beliefs, but it is a highly unlikely property to see vectors and linearity in the sense explained above, could the latter ultimately be \textit{derived} from the former? In other words, can the fact that we have vectorial addition of momenta, for example, ultimately be traced back to the linearity of probability theory?

In this paper, we argue that this speculation is not very bold after all, but in fact strongly supported by what we know about modern physics. We will present our line of argumentation in several steps. In Section~\ref{SecStatisticalModels}, we review and clarify how linearity appears as a crucial ingredient in \textit{statistical models}, in particular quantum theory, where two separate notions of linearity have to be distinguished. In Section~\ref{SecSpin}, we discuss that the very existence of \textit{spin} represents a link between two notions of linearity (probabilistic and spatial) that are a priori unrelated, in a way that is unprecedented in classical physics. This in itself is evidence for the hypothesis raised above. However, we know that low-energy quantum mechanics is only an approximation to a more fundamental theory, presumably quantum gravity. We therefore analyze in Section~\ref{SecRenormalization} what quantum gravity can tell us about the linear structure of quantum mechanics in our regime. We argue that \textit{independently} of the specific form of the fundamental theory, low-energy quantum mechanics arises from that theory via some sort of renormalization procedure which preserves linearity by construction. We thus conclude that physics as we know it supports our hypothesis, and discuss in Section~\ref{SecConclusions} how our line of argumentation fits into the widespread paradigm of emergent spacetime.

\section{Linearity in statistical models and quantum theory}
\label{SecStatisticalModels}
In every possible world, agents face uncertainty, and it is probability theory which is our major tool to deal with this uncertainty in a scientific manner. {As physicists, we are usually interested in a particular formulation of probability theory that takes the empirical method, or the experiment, as its starting point.}
Following Holevo~\cite{Holevo}, we call the elements of this formulation \textit{statistical models}\footnote{Up to minor differences, this formalism has been reinvented many times under many different names. It permeates quantum information theory under the names of ``generalized probabilistic theories''~\cite{Barrett} and ``convex operational theories''~\cite{BarnumWilce}; much earlier, the notion of a generalized state space has been mathematically formulated in terms of ``complete base norm spaces'', see e.g.~\cite{Davies}.}, and we will see that they encompass in particular quantum theory (and in fact generalize it). Consider any experiment that we would like to perform, which may or may not involve physics at some microscopic scale (what we are saying could, at least at the current stage, also apply to other fields of science). Suppose that we can make the following assumption~\cite{Holevo}: \textit{``The individual results in a sequence of identical, independent realizations of an experiment may vary, but the occurrence of one or another result in a long enough sequence of realizations can be characterized by a definite stable frequency.''}\footnote{Here we follow Holevo's exposition~\cite{Holevo} which uses frequentist terminology. However, it is perfectly possible to interpret statistical models in a Bayesian way, and to think of a ``state'' as a description of an agent's beliefs. In particular, interpretations of this kind for quantum theory have been advocated, e.g., in ``Relational Quantum Mechanics"~\cite{RQM} and in ``QBism''~\cite{QBism}. In fact, the formalism of quantum theory can even be reconstructed from a Bayesian perspective and operational rules that restrict an agent's acquisition of information from the observed systems~\cite{Hoehn2014,Hoehn2015,Hoehn2016}. This manifests the consistency of such an informational interpretation of quantum theory (and specifically of the quantum state as a ``catalogue of knowledge''). Note also that the formalism of statistical models does not at all commit us to an instrumentalist account of physics: a preparation procedure could as well be, say, the emission of a photon in a distant galaxy.} To avoid technicalities, let us for now only consider experiments with a discrete, finite number of possible outcomes, then the result of the experiment (obtained from many repetitions) corresponds to a discrete probability distribution over those outcomes.

\begin{figure}[!hbt]
\begin{center}
\includegraphics[angle=0, width=.5\textwidth]{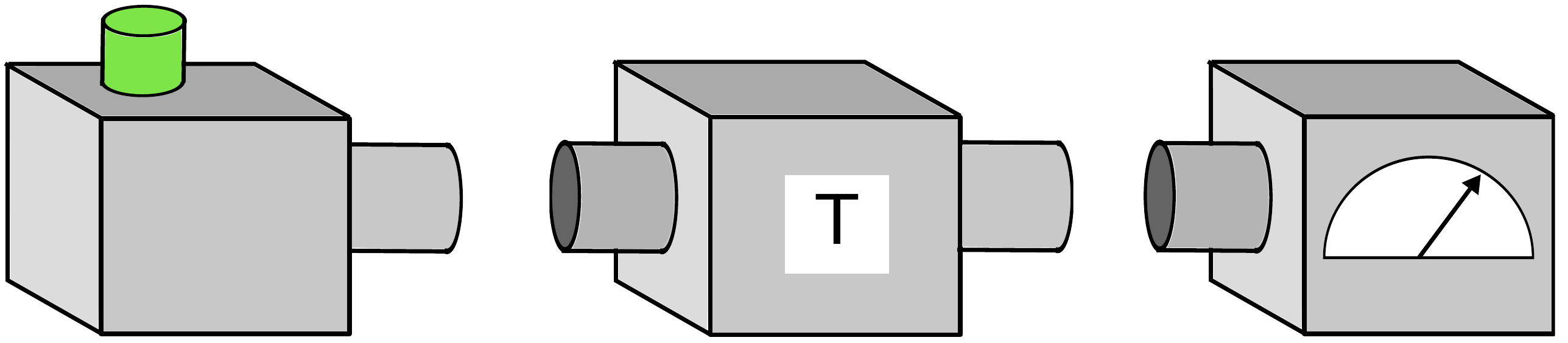}
\caption{A schematic illustration of the notion of experiment that is used to motivate the definition of a statistical model. Three stages of the experiment are distinguished: preparation, transformation, and measurement. Note that the applicability of the formalism of statistical models is not restricted to this specific simple situation.}
\label{fig_ptm}
\end{center}
\end{figure}

Figure~\ref{fig_ptm} illustrates that we can think of an experiment as consisting of (at least) two stages~\cite{Holevo}: \textit{``At the first stage of \textbf{preparation} a definite experimental set-up is settled, some initial conditions or `input data' of the experiment are established. At the following stage of the experiment the `prepared' object is coupled to a measuring device, resulting in [...] output data.''} The preparation is schematically illustrated in Figure~\ref{fig_ptm} on the left as a device which, after pressing a button, produces a certain physical system (say, a particle); the process which yields a measurement outcome is depicted as a device on the right. In many cases, we may want to separately consider a further part of the experiment, namely a ``transformation'' $T$ which summarizes whatever happens in between preparation and measurement. We are always free to regard the transformation as part of the preparation (``Schr\"odinger picture'') or as part of the measurement (``Heisenberg picture''), and thus recover Holevo's two-stage description.
A simple physical example is given by a Stern-Gerlach experiment: first, a silver atom is prepared in a certain quantum state; then a magnet acts as a transformation device, deflecting the atom depending on the spin of one of its electrons; and finally, a photographic plate acts as a position measurement.

The most fundamental concept of the mathematical description of a statistical model is that of a ``state''. Consider the class of all available preparation and measurement procedures. Suppose that two preparation procedures $P_1$ and $P_2$ have the property that they yield exactly the same statistics for all measurement procedures; in other words, no measurements can distinguish the physical systems that they prepare. In this case, we call $P_1$ and $P_2$ \textit{equivalent}. A \textbf{state} $\omega$ is then an equivalence class of preparation procedures.

Let us denote a measurement by $M$, and label the possible outcomes by $m=1,2,3,\ldots,N$, where $N$ is the finite number of possible outcomes of the measurement $M$. Denote the probability of obtaining outcome $m$ if performing measurement $M$ on state $\omega$ by ${\rm Prob}(m|\omega,M)$. According to our definition above, two states $\omega_1,\omega_2$ are identical if and only if they give the same outcome probabilities on all measurements:
\[
   \omega_1=\omega_2\quad\Leftrightarrow\quad {\rm Prob}(m|\omega_1,M)={\rm Prob}(m|\omega_2,M)\mbox{ for all }m\mbox{ and for all }M.
\]
In other words, a state is nothing but a ``catalog of probabilities'' for all possible measurements. Following Holevo (while slightly modifying his notation), we can then introduce a notion of ``mixed state''~\cite{Holevo}: \textit{``Let $\{\omega_i\}$ be a finite collection of states. Consider an infinite series of individual experiments in each of which the object is prepared in some of the states $\omega_i$, the occurrence of different values of $i$ being characterized by a probability distribution $\{p_i\}$. Denoting such `mixed' state by} $\omega=(\{\omega_i\},\{p_i\})$,
\textit{we have for any measurement $M$}
\begin{equation}
   {\rm Prob}(m|\omega,M)=\sum_i p_i {\rm Prob}(m|\omega_i,M).
   \label{eqMixture}
\end{equation}
\textit{Thus, we are led to adopt that for any finite set of states $\{\omega_i\}$ and any probability distribution $\{p_i\}$ there is a uniquely defined `mixed' state $\omega$ which is characterized by~(\ref{eqMixture}).''}

\begin{figure}[!hbt]
\begin{center}
\includegraphics[angle=0, width=.5\textwidth]{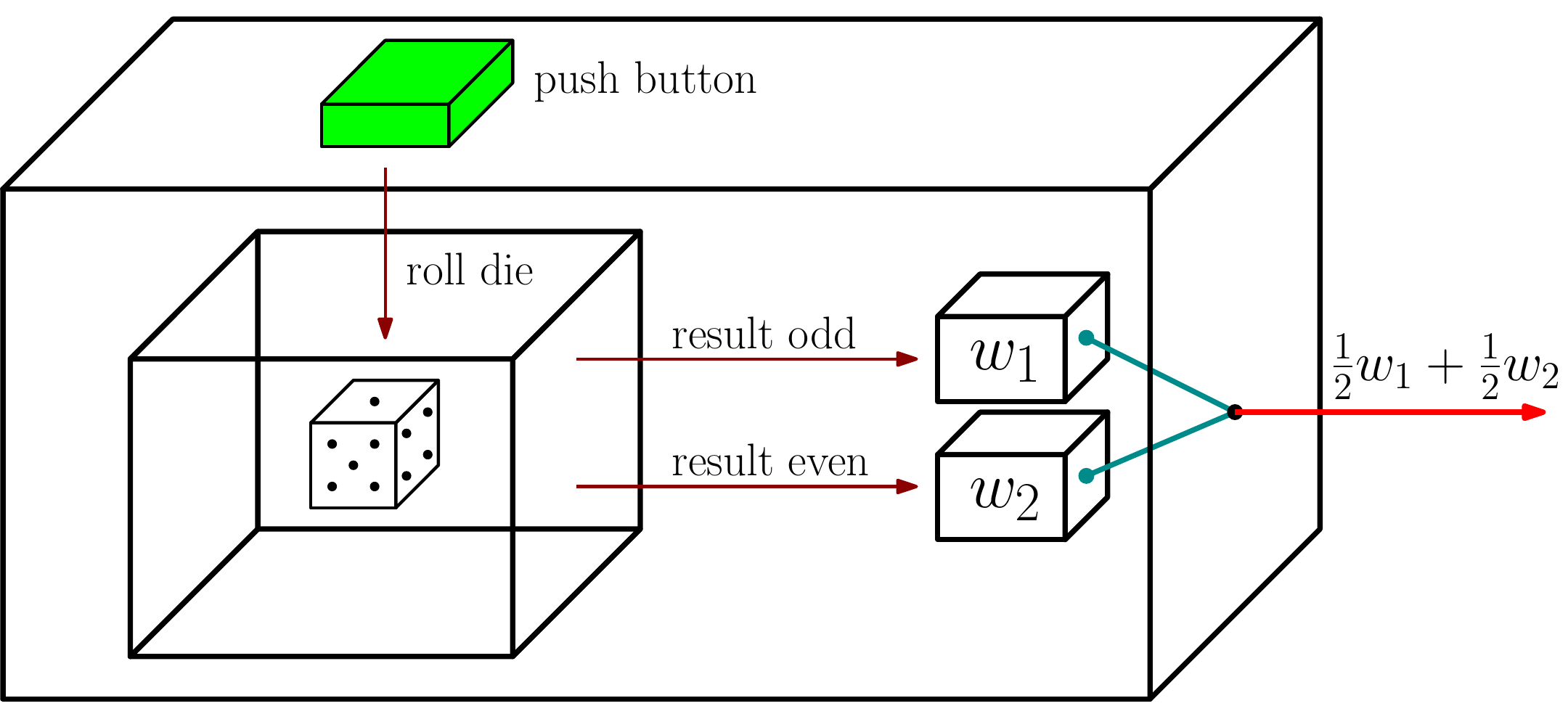}
\caption{An illustration of the general idea of a mixed state as explained in the main text.}
\label{fig_mixed}
\end{center}
\end{figure}

Figure~\ref{fig_mixed} shows a colourful illustration of this idea: a device prepares either state $\omega_1$ or state $\omega_2$ with probability $1/2$ each, resulting in the preparation of a mixed state $(\{\omega_1,\omega_2\},\{\frac 1 2,\frac 1 2 \})$. The figure already anticipates that we can write this mixed state as $\frac 1 2 \omega_1+\frac 1 2 \omega_2$. This is true due to the following theorem\footnote{In more detail, the definition of states as equivalence classes and the characterization of mixed states via~(\ref{eqMixture}) turn the set of states into a \textit{separated mixture space} in the sense of~\cite{Holevo}, and so the first half of Theorem~\ref{TheConvex} follows from Proposition 2.1 in~\cite{Holevo}. For the second half, note that measurement outcome probabilities are affine functionals on that mixture space, which are also linear functionals on the ``bidual'' space $A:=\mathfrak{L}$ that is constructed in the proof of Proposition 2.1.}, which follows from Proposition 2.1 in~\cite{Holevo}:
\begin{theorem}
\label{TheConvex}
We can always represent the states $\omega$ as elements of some linear space $A$, such that mixed states are represented as
\[
   (\{\omega_i\},\{p_i\})=\sum_i p_i\omega_i.
\]
Moreover, every outcome $m$ of every measurement $M$ is represented by a linear functional $e_m^{(M)}$ on $A$ such that
\[
   {\rm Prob}(m|\omega,M)=e_m^{(M)}(\omega).
\]
\end{theorem}
Thus, statistical models have a natural notion of linearity that comes from the linearity of statistical mixing. As a simple add-on, we can also introduce a notion of ``normalization'' (the total probability to obtain an outcome in any measurement), which is a linear functional $u_A$ that attains the value $1$ on any state. To avoid technicalities, let us restrict to the case of finite-dimensional statistical models, then we obtain the following definition:\footnote{We are ignoring here some details that are often discussed in the framework of generalized probabilistic theories. For example, for any given effect $e$, it is not a priori clear whether that effect can actually be implemented in a physical measurement. In other words, there may be a subset of ``physically allowed effects''; by calling a functional ``effect'' we are \textit{not} claiming that it can actually be physically implemented. Similar reasoning holds for transformations. However, for the purpose of this paper, this distinction is mostly irrelevant. Furthermore, we are ignoring the notion of ``equivalence'' of statistical models, which basically says that the linear reparametrization of a statistical model is physically indistinguishable from the original statistical model~\cite{MM3D}. This means in particular that one should define the notion of a ``quantum model'' more carefully by allowing arbitrary linear reparametrizations.}
\begin{definition}
\label{DefStatisticalModel}
A \textup{statistical model} is a triple $(A,u_A,\Omega_A)$, where $A$ is a finite-dimensional real vector space, $u_A:A\to\R$ is a non-zero linear functional on $A$, and $\Omega_A$ is a compact\footnote{From the considerations above, one can motivate that state spaces must be bounded. This is because states are uniquely characterized by measurement outcomes probabilities, i.e.\ effects which are (bounded) functionals. For mathematical simplicity, we assume that state spaces must also be topologically closed (for a physical motivation, see e.g.~\cite{MasanesMueller}).} convex subset of full dimension in the hyperplane of those $a\in A$ where $u_A(a)=1$.

An \textup{effect} $e$ is a linear functional on $A$ such that $0\leq e(\omega)\leq 1$ for all $\omega\in\Omega_A$. Effects describe probabilities of outcomes of measurements. A (normalization-preserving) \textup{transformation} is a linear map $T:A\to A$ such that $T(\Omega_A)\subset \Omega_A$ (i.e.\ states are mapped to states). It is \textup{reversible} if and only if $T^{-1}$ exists and is a transformation, too.
\end{definition}
It is important to understand why transformations must be linear. To this end, consider the device in Figure~\ref{fig_mixed}, and substitute it as a preparation device in the scenario of Figure~\ref{fig_ptm}. Consider the mixed state $\omega=\frac 1 2 \omega_1+\frac 1 2 \omega_2$. Clearly, what enters the measurement device is, by definition, the state $T\omega$. On the other hand, we can consider the union of the preparation and measurement device as a new (joint) preparation device, which prepares the state $T\omega_1$ with probability $1/2$, and the state $T\omega_2$ with probability $1/2$. Consistency of both viewpoints demands that $T\left(\frac 1 2 \omega_1+\frac 1 2\omega_2\right)=\frac 1 2 T\omega_1+\frac 1 2 T\omega_2$, and similarly for other mixtures. From this one can show that $T$ corresponds uniquely to a linear map on $A$~\cite{Hardy2001,Barrett}.

Classical probability distributions $\omega=(p_1,\ldots,p_n)$ with $p_i\geq 0,\sum_i p_i=1$ can be seen as states in a \textit{classical statistical model} where $A=\R^n$, $u_A(x)=x_1+\ldots+x_n$ and $\Omega_A$ the set of all discrete probability distributions. Nonclassical statistical models can arise in a variety of situations. For example, Holevo~\cite{Holevo} has shown that they can arise if the set of measurements that can be performed on a classical statistical model is limited for some reason.\footnote{In fact, \textit{all} statistical models can be obtained in this way, cf.~\cite{Holevo}. This is related to the problem of hidden variables.} A nice example (though in a different formalism) is given by Spekkens' toy model~\cite{Spekkens}. In particular, \textbf{quantum theory} can be regarded as a special case of statistical models:
\begin{definition}[Quantum model]
\label{DefQuantumModel}
For every $n\in\mathbb{N}$, the $n$-level \textup{quantum model} is defined as the statistical model $(A_n,u_n,\Omega_n)$, where $A_n$ is the real linear space of Hermitian complex $n\times n$ matrices, $u_n(\rho):=\tr(\rho)$ is the trace functional, and $\Omega_n\subset A_n$ is the set of $n\times n$ density matrices.
\end{definition}
Many ingredients of operational quantum theory follow directly from Definition~\ref{DefStatisticalModel}. For example, it follows from elementary linear algebra that the effects are exactly the maps $\rho\mapsto \tr(\rho E)$, where $0\leq E \leq\mathbf{1}$, and thus every measurement is described by a positive operator-valued measure (POVM). In other words, the (density matrix version of the) Born rule follows trivially.

Furthermore, Wigner's Theorem gives us the reversible transformations as a mathematical consequence:
\begin{theorem}[Wigner's Theorem~\cite{Bargmann}]
Every reversible transformation $T$ on a quantum model is of the form
\[
   T(\rho) = U\rho U^{-1},
\]
where $U$ is either a unitary or an antiunitary map.
\end{theorem}
It turns out that it is physically impossible to implement conjugations by antiunitary maps. This empirical fact can be theoretically explained by either one of the two following arguments: first, by the constraint that it must be possible to implement every transformation $T$ continuously in time, and hence $T$ must be in the connected component of the identity map (which the antiunitary maps are not). Second, and more importantly, one can also argue that applying antiunitary conjugations to one half of an entangled state yields negative probabilities~\cite{NielsenChuang}.

Thus, we are left with the unitary conjugations. This gives us the usual linearity in the state vector in a second step: since all pure states\footnote{In a general statistical model, a state $\omega\in\Omega$ is called \textit{pure} if it cannot be written as a non-trivial convex combination of other states.} are of the form $\rho=|\psi\rangle\langle\psi|$, we can formally take the map $|\psi\rangle\langle\psi|\mapsto U|\psi\rangle\langle \psi|U^\dagger$, and ``split'' it into two halves (up to a complex phase). This gives us the usual notion of unitarity, $|\psi\rangle\mapsto U|\psi\rangle$, and thus the superposition principle. We obtain two different notions of linearity (probabilistic linearity versus state vector linearity) that have substantially different status:

\begin{center}
\includegraphics[angle=0, width=.8\textwidth]{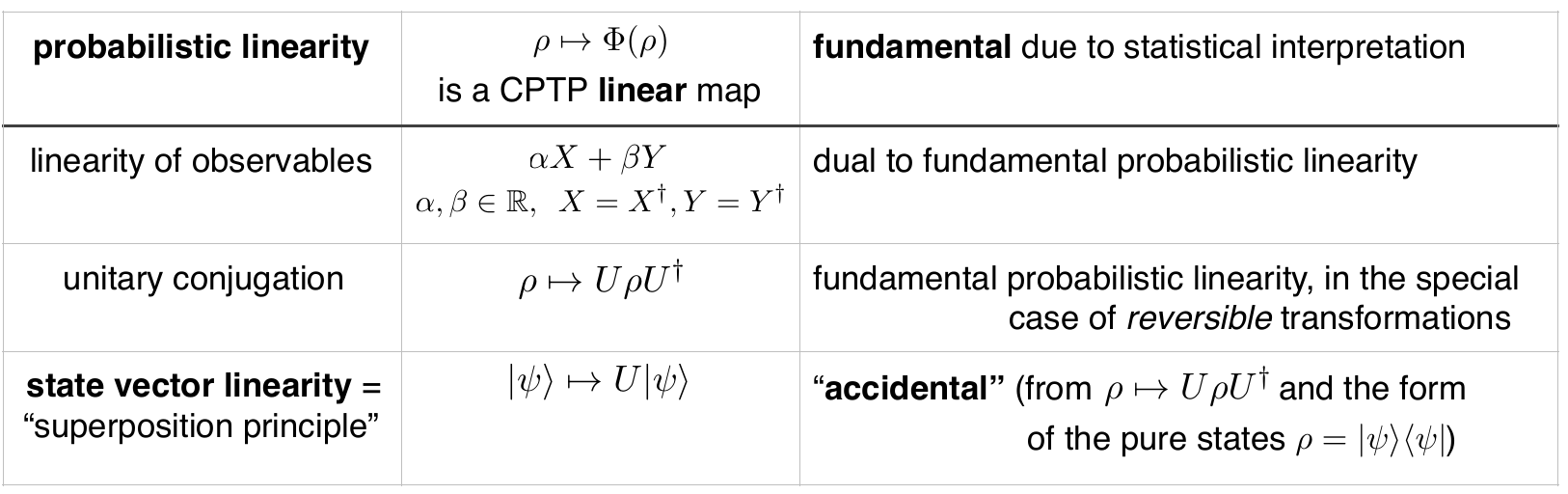}
\label{fig_table}
\end{center}

The fact that arbitrary, not necessarily reversible transformations on quantum states (for example, Markovian time evolution of open systems) are given by completely positive, trace-preserving (CPTP) linear maps follows again from considering the action of transformations on composite systems and entangled states~\cite{NielsenChuang}. (Our definition of statistical models above does not incorporate any notion of composite systems, but this can be included without much effort~\cite{Barrett}).

We can define a general notion of \textit{observables} in statistical models as those objects $X$ that yield real expectation values $X(\omega)\in\R$ on states $\omega$. Due to the interpretation of mixed states above, we demand that observables (being \textit{expectation values}) satisfy an analog of equation~(\ref{eqMixture}), i.e.\ are \textit{linear} functionals\footnote{Following the terminology of Holevo~\cite{Holevo} in more detail, we would first postulate that observables are affine functionals on the separated mixture space, and then, by the same construction as for effects, they become linear functionals on $A$.} on the space of states, $A$. In other words, an observable $X$ is an element of the dual space, $X\in A^*$. Effects are special observables, and they span the linear space $A^*$. Thus, every observable $X\in A^*$ can be written (in general in many different ways) as a linear combination $X=\sum_{i=1}^n \lambda_i e_i$ with effects $e_1,\ldots,e_n$ such that $\sum_i e_i=u_A$ and real numbers $\lambda_1,\ldots,\lambda_n$. This gives us a physical interpretation of observables by telling us how to measure them: perform the $n$-outcome measurement described by the $n$ effects $e_1,\ldots,e_n$, and assign the value $\lambda_i$ to outcome $e_i$. Repeat the measurement many times, and determine the average of the obtained values in the long run.

Since $A^{**}=A$ in finite dimensions, this tells us that we can also interpret states as linear functionals on the space of observables, which is the more standard view in mathematical physics~\cite{BratteliRobinson}.\footnote{Note that observables in statistical models lack two pieces of structure that they have in the special case of quantum theory: first, there is in general no notion of a distinguished ``projective measurement'' with outcomes corresponding to the ``eigenvalues'' of a given observables; and second, observables cannot be multiplied, i.e.\ there is no notion of algebra. Clearly, the question arises how we assign values $\lambda_i$ to the different outcomes. The answer will depend on the physical context; in particular, in those special situations where we have post-measurement states and repeatability of measurements (as in quantum theory), these outcome values will sometimes characterize conserved quantities, carried by the post-measurement quantum systems, that can be compared across different quantum systems, yielding an operational notion of ``scale''.} However, having probabilities as the starting point, it is more natural for us to take the reverse point of view and consider observables as the derived notion, dual to the states. Therefore, if we calculate a linear combination of observables like $\alpha X+\beta Y$, then this linear structure is inherited from the fundamental probabilistic linear structure of the state space.

The lesson to draw from the construction above is thus contrary to what is usually taught in introductory quantum mechanics textbooks: it is not the superposition principle that is the fundamental principle of quantum mechanics, but it is rather the structure of the set of states as given in Definition~\ref{DefQuantumModel}. Linearity in the state vector is an accidental consequence of this structure and of the fundamental linearity of probabilities.

Even though this point of view seems surprising to some working physicists at first sight, it is actually folklore in quantum information theory~\cite{NielsenChuang} and algebraic formulations of quantum (field) theory, where states are defined as positive linear functionals on observables~\cite{BratteliRobinson}. From a less formal starting point, this insight is most clearly demonstrated by the recent results on ``reconstructions'' of quantum mechanics. These works show that quantum theory, as specified in Definition~\ref{DefQuantumModel}, can be derived from simple information-theoretic principles~\cite{Fuchs,Hardy2001,DakicBrukner,MasanesMueller,Chiribella,Hardy2011,Masanes2013,Barnum2014,Hoehn2014,Hoehn2015,Wilce2016}\footnote{There has been a long prehistory of earlier attempts along similar lines, starting with Birkhoff and von Neumann~\cite{Birkhoff}, including a long history of research in quantum logic and some earlier attempts of axiomatization like e.g.\ Ludwig's~\cite{Ludwig}. The newer reconstructions benefit from many insights of quantum information theory, and from a finite-dimensional setup that may have been considered less natural before the advent of quantum information theory.}, analogously to the way that the Lorentz transformations can be obtained from the relativity principle and the invariance of the speed of light. They start with a general framework, more or less equivalent to the statistical models framework above, give a small set of well-motivated postulates, and then show that quantum models (and, in some cases, classical models) are picked out as the unique solutions. For example, the postulates in~\cite{MasanesMueller} are essentially\footnote{We are dropping here the first postulate (finite-dimensionality of the generalized bit) because it is usually assumed without stating it; and we are also dropping the fifth postulate (``all mathematically well-defined measurements on a generalized bit are also physically allowed'') -- in order to even state this postulate, one needs to introduce the additional structure of ``being physically allowed'' into the formulation which complicates the setup.}:
\begin{enumerate}
	\item \textbf{Tomographic locality:} The state of a composite system is characterized by the statistics of the measurements on the individual components.
	\item \textbf{Subspace postulate:} All systems that effectively carry the same amount of information have equivalent state spaces.
	\item \textbf{Symmetry:} Any pure state of a system can be reversibly transformed into any other.
\end{enumerate}
A different set of postulates, which is inspired by Rovelli's relational quantum mechanics~\cite{RQM} and the Brukner-Zeilinger informational interpretation \cite{zeilinger1999foundational,Brukner:ys,Brukner:vn}, can be found in~\cite{Hoehn2014,Hoehn2015}, see also the review~\cite{Hoehn2016}; it has the virtue of starting from a manifestly Bayesian perspective, thereby putting an agent's acquisition of information from the observed systems centre stage. The postulates of the various reconstructions turn out to have quantum (and, in some cases, also classical) models as their unique solutions.
Crucially, in these reconstructions, it is the \textit{set of density matrices} that is obtained first, not the notion of a state vector. In other words, the first linear space that appears in these reconstruction is the \textit{real-linear space of Hermitian matrices} over $\mathbb{C}^n$, \textit{not} the complex Hilbert space $\mathbb{C}^n$ itself. For example, it is usually a first step to prove that a generalized bit (i.e.\ two-level system) must be described by a state space that is a Euclidean ball (like the Bloch ball of a quantum bit), and only after this one uses the fact that elements of this ball can be represented by $2\times 2$ density matrices. This supports the classification of density matrix linearity as fundamental and state vector linearity as a derived concept.

\section{Relating two kinds of linearity: spin}
\label{SecSpin}
In our physical world, there is yet another kind of linear structure, namely that of the local tangent space of our spacetime manifold.
For the moment, let us focus on a situation where all involved physical systems move slowly enough so that we can ignore relativistic effects (we will later lift this restriction). Furthermore, we only consider small regions of spacetime where gravity plays no significant role. In this regime, we can talk about Galilean space as in classical mechanics, and we have a notion of (three-)vectors that describe quantities like position or velocity. To emphasize that we are talking about a kind of linearity that has a completely different physical interpretation than in the previous section, we denote linear combinations by different symbols. If $\vec v,\vec w$ are in the tangent space $V$, and $\lambda,\mu\in\R$, then we write
\[
   (\lambda\odot \vec v) \oplus (\mu \odot\vec w) \quad \in V
\]
for the corresponding linear combination, and we will also use ``$\odot$'' for matrix multiplication.

A priori, this kind of linearity has nothing to do with the linearity of statistical mixtures from Section~\ref{SecStatisticalModels}. However, quantum mechanics directly relates these two notions of linearity. The most transparent instance of this can be seen by example of the spin-$1/2$ particle. From a mathematical point of view, the origin of this relation is the fact that we have a (projective) representation of the rotation group\footnote{The terminology of ``representation'' itself carries an implicit physical assumption, namely that it is $R$ that is fundamental (representing a space-time symmetry), and that this fundamental notion is mathematically carried over to (i.e.\ represented on) the Hilbert space. In the following we argue that one should instead reverse the perspective: perhaps it is the Hilbert space and its symmetries that is more fundamental (following directly from a statistical model), and spacetime symmetries are in 
some way inherited from it.}, which assigns to every rotation $R\in {\rm SO}(3)$ a corresponding unitary $U_R\in{\rm SU}(2)$. Since the actually significant operation is conjugation by unitaries, $\rho\mapsto U_R \rho U_R^\dagger$, what is really of interest is the corresponding element $U_R\cdot U_R^\dagger$ of the projective unitary group ${\rm PU}(2)$. Now we have a one-to-one correspondence ${\rm SO}(3)\leftrightarrow {\rm PU}(2)$, where ${\rm SO}(3)$ is a group of linear transformations \textit{on tangent space}, whereas ${\rm PU}(2)$ are linear transformations acting \textit{on the linear space of qubit observables}. In other words, the spin-$1/2$ representation relates {the} linear structure on tangent space directly to {the} linear structure on the space of the (quantum) statistical model.

To explore this correspondence in a physical context, consider a standard Stern-Gerlach device. The electron interacts with the gradient of a non-homogeneous magnetic field,
\[
   \vec f = \vec\nabla |B| \in V,
\]
which determines the direction in which spin is measured. In general, for any tangent space vector $\vec f$, we have an observable $Z_{\vec f}$ that is measured in the Stern-Gerlach experiment if the (gradient of the) magnetic field is set to $\vec f$. For example, if we introduce coordinates in tangent space such that the first coordinate is defined to be the $z$-coordinate, we may have (for $\lambda\in\R$ and $\hbar\equiv 1$)
\begin{equation}
   Z_{\lambda\odot (1,0,0)} =\lambda \Delta\, \frac 1 2\left(\begin{array}{cc} 1 & 0 \\ 0 & -1
   \end{array}\right),
   \label{eqSGObservable}
\end{equation}
i.e.\ the Pauli-$Z$-matrix is implemented, times a factor (length scale) $\Delta>0$ such that this observable quantifies the \textit{deflection} of the particle after the magnet ($\Delta$ will be proportional to the strength of the magnetic field). In general,
\[
   Z_{\underbrace{R\odot \vec f}_{\mbox{spatial}}} = \underbrace{U_R Z_{\vec f} U_R^\dagger}_{\mbox{statistical}}
\]
which combines two separate notions of linearity in a single equation: a spatial-linear map, $\vec f \mapsto R\odot \vec f$, and a statistical-linear map, $Z_{\vec f} \mapsto U_R Z_{\vec f} U_R^\dagger$. The latter is directly related (by duality) to fundamental probabilistic linearity, as pointed out in Section~\ref{SecStatisticalModels}.
Operationally, this relation can be used to \textit{infer spatial-linear structure from statistical-linear structure} in the laboratory. To this end, consider the following thought experiment:\\

\textbf{Thought experiment 1:} Alice the physicist has bought three different Stern-Gerlach measurement devices from a manufacturer. The devices are characterized by magnetic field gradients $\vec f, \vec g, \vec h \in V$, and the manufacturer promises that these gradients satisfy $\vec h =\vec f \oplus \vec g$. In the factory, this is achieved by building three versions of a standard magnet, with suitable strengths and orientations, inside the devices. However, for patent protection reasons, the manufacturer has put the devices inside sealed black boxes, and Alice cannot directly verify this claim by inspecting the spatial orientation of the magnets. \textit{(Alternatively, Alice may simply lack the physical means to measure spatial orientations.)} All she can do is to input quantum states into the devices, and inspect the corresponding measurement results: the deflection in units of inches is shown on a digital display, but the particles do not leave the black box.

Despite the black box, it is possible for Alice to verify the manufacturer's claim: by tomography (i.e.\ preparing all kinds of known quantum states and inspecting the corresponding outcome statistics), she can determine the observables $Z_1,Z_2,Z_3$ that are measured by the three devices, and verify\footnote{For any given device (say, one that measures $Z_{\vec f}$), the two possible outcomes on the digital display will always be $\pm |\vec f|\Delta/2$, and so a necessary condition that Alice can check directly is whether $|\vec h|\leq |\vec f|+|\vec g|$.} whether $Z_3=Z_1+Z_2$ (corresponding to $Z_{\vec h}=Z_{\vec f}+Z_{\vec g}$). Note that the exact form of the observables $Z_i$ as matrices depend on her arbitrary choice of basis, but the identity $Z_3=Z_1+Z_2$ does not.\\

Typically, approaches to quantum gravity consider our notion of smooth spacetime manifold to be non-fundamental, emergent from some underlying microscopic quantum theory. In particular, this view suggests that the notion of (linear) tangent space is emergent too. But then the previous thought experiment suggests a well-informed speculation: it proposes that \textit{the linear structure of tangent space might be inherited from quantum theory's probabilistic linear structure} (more in detail, from the dual linear structure on the space of observables). After all, quantum theory is all that is needed in the thought experiment to determine the linear structure of space. Note that this is a mathematical statement: we have some form of mathematical structure in our world (spatial linearity), and we claim that this structure appears in the architecture of physics as a consequence of another structure (probabilities). How this comes about in detail is something that we make no claim about -- the answer to this question will rely on the specific characteristics of the microscopic theory (say, quantum gravity), and this theory will then also have to explain the origin of the group representation that we started with.

Since we have initially studied the example of spin-$1/2$ above, it may seem as if this value of spin was somehow special. However, this is not true: for any spin $s$, spatial rotations $R$ are in one-to-one correspondence to unitary conjugations $U_R\cdot U_R^\dagger$ on the corresponding Lie algebra of spin observables, which is a three-dimensional subspace of the vector space of observables (on which (the adjoint representation of) ${\rm SO}(3)$ acts). Therefore, we have a correspondence of spatial and statistical linear structure for every value of spin, and the thought experiment above can be run in all cases, with the same conclusion.

In fact, the notion of spin does not only allow us to infer the linear structure of space, but also its Euclidean structure. Consider the following variant of the thought experiment:\footnote{Another variant of thought experiment 2 has been given in~\cite{MM3D}.}\\

\textbf{Thought experiment 2:} Alice the physicist has bought two Stern-Gerlach measurement devices from the same manufacturer as in thought experiment 1. They are also hidden inside of black boxes, and they are even somewhat simpler: instead of the deflection, they only show ``$\oplus$'' or ``$\ominus$'' on the display after the measurement, depending on whether the spin was measured as up or down. Alice would like to know the angle between the devices' field gradients, $\angle (\vec f,\vec g)$, but she cannot inspect the magnets directly.

However, she can find the answer indirectly. First, by tomography, she can find the observables $Z_{\vec f}$ and $Z_{\vec g}$ (in some basis) that both devices are measuring, up to a global factor (corresponding to $\lambda\Delta$ in~(\ref{eqSGObservable})), because she does not know the deflection or strength of the magnetic field. Then it is easy to check (e.g.\ in the Zeeman basis) that
\[
   \langle Z_{\vec f},Z_{\vec g}\rangle_{\rm HS} \equiv \tr(Z_{\vec f}Z_{\vec g})=\alpha_s \Delta^2 \, \vec f \cdot \vec g,
\]
where $\alpha_s=s(s+1)(2s+1)/3$ is a constant that depends only on the spin quantum number $s$ of the particle, and $\langle X,Y\rangle=\tr(XY)$ is the Hilbert-Schmidt inner product on the real vector space of observables, with corresponding norm $\|X\|_{\rm HS}=\sqrt{\langle X,X\rangle_{\rm HS}}$. Thus
\[
   \angle (\vec f,\vec g)=\arccos \frac{\vec f \cdot \vec g}{\|\vec f\|\cdot \|\vec g\|}=\arccos\frac{\langle Z_{\vec f},Z_{\vec g}\rangle_{\rm HS}}{\|Z_{\vec f}\|_{\rm HS}\cdot \|Z_{\vec g}\|_{\rm HS}}
\]
which Alice can use to determine the angle from tomography, i.e.\ probability measurements only.

Therefore, not only is it consistent to regard the linear structure of tangent space as inherited from quantum theory, but also its inner product structure which allows to compute lengths and angles. We have a correspondence
\[
   \langle Z_{\vec f},Z_{\vec g}\rangle_{\rm HS}\mbox{ inner product on the vector space of observables } \leftrightarrow \vec f \cdot \vec g\mbox{ inner product on tangent vector space}.
\]
Representing observables $X$ as self-adjoint matrices means that we compute expectation values on states $\rho$ via
\[
   \langle X\rangle_{\rho}=\tr(\rho X)=\langle \rho,X\rangle_{\rm HS},
\]
and so, by duality, we can apply the Hilbert-Schmidt inner product also to states, i.e.\ density matrices. But is there any \textit{information-theoretic} significance to this inner product? In the case of \textit{linearity}, we have seen in Section~\ref{SecStatisticalModels} that there is a direct operational interpretation, namely statistical mixing. And indeed, it turns out that the Hilbert-Schmidt inner product has an information-theoretic origin too. The clearest way to see this is again in the general context of statistical models, where we have the following result~\cite{MuellerUdudec}.

\begin{theorem}[\cite{MuellerUdudec}]
\label{TheBitSymmetry}
Consider any statistical model with the following property (``bit symmetry''): if $\omega_1,\omega_2$ are perfectly distinguishable pure states, and so are $\varphi_1,\varphi_2$, then there is a reversible transformation $T$ such that $T\omega_1=\varphi_1$ and $T\omega_2=\varphi_2$.

If a statistical model $(A,u_A,\Omega_A)$ has this property, then there is an inner product $\langle\cdot,\cdot\rangle$ on $A$ that is non-negative on all states, satisfies $\langle\omega,\omega\rangle=1$ for all pure states $\omega$, $\langle\varphi,\varphi\rangle<1$ for all mixed states $\varphi$, $\langle\omega,\varphi\rangle=0$ if $\omega$ and $\varphi$ are perfectly distinguishable, and $\langle T\omega,T\varphi\rangle=\langle \omega,\varphi\rangle$ for all reversible transformations $T$. Moreover, identifying $A^*$ and $A$ via this inner product, the set of unnormalized states and the set of unnormalized effects become identical (``self-duality'').
\end{theorem}
Indeed, quantum models have the property of bit symmetry: in quantum theory, pure states are perfectly distinguishable if they are orthogonal. But if $\omega_1,\omega_2$ are orthogonal, and so are $\varphi_1,\varphi_2$, then there is some unitary that maps one pair to the other. Thus Theorem~\ref{TheBitSymmetry} guarantees that there is an inner product on the vector space of Hermitian matrices which has all the claimed properties. And this is exactly the Hilbert-Schmidt inner product (the last property above is $\langle U\rho U^\dagger, U\sigma U^\dagger\rangle_{\rm HS}=\langle \rho,\sigma\rangle_{\rm HS}$).
Note that this theorem gives directly an inner product on the space of states (or observables), i.e.\ the linear space of Hermitian matrices, and \textit{not} on Hilbert space itself. This is reminiscent of the insights of Section~\ref{SecStatisticalModels}, where Hilbert space and state vectors turned out to be derived concepts as well. In contrast, the standard mathematical postulates of quantum mechanics would start with a Hilbert space, and characterize the Hilbert-Schmidt inner product by linear extension of
   $\langle |\varphi\rangle\langle\varphi|,|\psi\rangle\langle\psi|\rangle_{\rm HS}=|\langle\varphi|\psi\rangle|^2$,
where the inner product on the right-hand side is the inner product of the underlying Hilbert space. This is not the path taken here, but it will become important in the next section.

In a nutshell, the inner product comes from the \textit{symmetry} of quantum theory's state space: there are ``many'' transformations $T$ as in Figure~\ref{fig_ptm} that preserve information in the sense that they can in principle be reversed (see, e.g., \cite{Hoehn2014}), which implies that there is such an inner product. As shown in the thought experiment above, this corresponds (via group representation) to the Euclidean inner product on the tangent vector space.

So far, we have only considered non-relativistic space -- do the insights from above carry over to the relativistic case? Insofar as Euclidean space is a special case, or ``subspace'', of relativistic space-time, our arguments above establish a relation between statistical linearity and spatial linearity for at least part of the relativistic structure. Analyzing this relationship in more detail in the relativistic case would amount to study technically more difficult questions, involving relativistic spin operators or infinite-dimensional Hilbert spaces (there are no non-trivial finite-dimensional unitary representations of the Lorentz group), which we will not do in this paper. However, our results \cite{Hoehn} show that one can establish analogous arguments in the relativistic context and, in particular, that one can consistently treat relativistic Stern-Gerlach devices in a similar manner (using some results from~\cite{Palmer}).

\section{Quantum gravity and renormalization}
\label{SecRenormalization}
Ultimately, our low energy observables (in particular the spin observables of Thought Experiments 1 and 2) will have to be understood as a ramification of a deeper layer of physical reality, presumably quantum gravity. These must include the quantum fields living on our background classical space-time, but also the geometric properties of the background itself. Hence, the proposal put forward in this article is most naturally understood as a conjecture about how the linearity of the fundamental observables of quantum gravity relates to the local linearity of space-time. The purpose of this section is to discuss how renormalization is expected to bridge the gap between the natural scale of quantum gravity -- presumably the Planck scale -- and the scale at which low-energy quantum mechanics operates. Crucially for our proposal, we will argue that the linearity of quantum theory itself will be preserved by such a coarse-graining procedure. Combining this with the arguments of Section~\ref{SecSpin}, where we relate the linearity of low-energy quantum mechanics to the linearity of tangent space via spin, finally established a bridge from the most fundamental quantum theory to space-time.

Obviously, since the basic architecture of quantum gravity is still largely debated, we may only speculate as to what might happen. In particular, a full solution to the quantum gravity conundrum might very well require a deep revision of the structure of quantum theory itself. For the purpose of the argument, we will refrain from diving into such uncharted territory and assume instead that the basic structure of quantum theory as we know it is maintained in the quantum gravity regime. That is, we postulate that quantum gravity is described by an algebra of (microscopic) observables which can be unitarily represented on a Hilbert space, and can, in principle, be measured by an external agent. 

This is a strong assumption, but we may expect that it will, at the very least, hold in some corner of any full quantum theory of gravity. Think for instance of an asymptotic agent looking at the sky in a semi-classical region of space-time (which may be modelled, e.g., by a space-time conformal boundary, or a sufficiently classical cosmological geometry at sufficiently late Hubble times), and detecting signals coming from a {spatiotemporally} distant region in quantum turmoil (e.g.\ a black hole forming and evaporating in the bulk, or some violent event in the very early universe). This assumption would also offer an asymptotic operational interpretation of the probabilistic structure underlying a quantum theory of gravity; in what follows, the probabilities encoded in the Hilbert space structure may be interpreted as those ``experienced'' by such an asymptotic observer, thereby sidestepping interpretational issues related to a ``wave function of the universe''. We emphasize that for the argument it will be irrelevant whether probabilities are interpreted in a frequentist, Bayesian or any other consistent manner. From an operational/experimental point of view, this idealized situation of asymptotic observers may actually encompass {most of what} there is to be said about quantum gravity.

Having assumed that quantum gravity should be formulated in the usual language of quantum theory, we may now invoke the discussion of Section~\ref{SecStatisticalModels} to argue that the fundamental linearity of quantum gravity states should be the unique origin of the two seemingly different linear structures that we observe at lower energy: the linearity of quantum field observables on the one hand; and, on the other hand, the linearity of {the} local space-time {structure} and other vectorial observables such as, say, momentum. 
Indeed, any quantum theory of gravity must reduce in a semi-classical regime to quantum field theory on a classical background space-time, and it is expected that this will be achieved by a suitable coarse-graining or renormalization of the fundamental quantum gravity dynamics. This implies that not only the effective quantum fields but also the geometric observables associated to the background will have to be realized as a coarse--graining of the microscopic quantum gravity observables.

Renormalization and coarse--graining are the fundamental tools which allow us to extract the large-scale collective properties of any complex quantum system with many degrees of freedom, including its emergent classical properties. In particular, there exist simple quantum-mechanical models showing that just going to `large quantum numbers' is generally not sufficient in order to derive classical behaviour of quantum systems with many degrees of freedom, and that rather some notion of coarse-graining is required as well \cite{kofler1,kofler2}. In fact, this is precisely how macroscopic and classical properties of quantum many-body systems are studied (see, e.g.~\cite{Vidal}). Thus, it is not unreasonable to anticipate that any explanation of classical properties of space-time from a deeper quantum gravity theory will presumably require a suitable coarse-graining and renormalization scheme. In this picture, classical space-time geometry emerges as the collective properties of underlying microscopic degrees of freedom. This view is also supported by a recent derivation of the (semiclassical) Einstein field equations from entanglement equilibrium conditions on a quantum field in space-time which hints at the necessity of a suitable coarse-graining and renormalization scheme for bridging the deep quantum gravity regime to the known physics \cite{ted}.\footnote{More precisely, Jacobson's
derivation~\cite{ted} separates the variation of the field's entanglement entropy into an ultraviolet (UV) and infrared part. The latter is determined by the usual von Neumann entropy of the field and yields the expectation value of the energy. By contrast, the UV part is state independent and scales with the area of the region based on the assumption of universal UV entanglement properties of vacuum-like states. This part gives rise to the Einstein tensor. The derivation suggests that one only obtains semiclassical general relativity if the UV physics has been appropriately coarse-grained to universal low-energy effects.}

But does renormalization preserve the linear structure of quantum theory? The short answer is 'yes', and it is so essentially by construction. Indeed, since renormalization operates within standard quantum theory, this means it must preserve its basic architecture, including its linear structure.\footnote{There exist additional operational plausibility arguments for the linearity-preserving property of renormalization. Assume (a) that quantum theory is universal, holding at all scales, and (b) that macroscopic observables, states and transformations of low-energy effective quantum theories also have a microscopic representation that permits their evaluation in a more fundamental quantum theory which is a pre-image under renormalization of the low-energy theory. Then any macroscopic quantum operation or measurement of a macroscopic observable must have a linear representation in the microscopic theory. If renormalization did not preserve linearity, the same operation or measurement would be represented with non-linear modifications in the low-energy effective quantum theory -- in conflict with experiment. Indeed, it is known that non-linear modifications of quantum theory lead to all kinds of pathological features (such as instantaneous signalling~\cite{Gisin1990,Simon2001} or a violation of the second law of thermodynamics~\cite{PeresSecondLaw} etc.) which are experimentally ruled out at low energy. As argued in Section~\ref{SecStatisticalModels} above, linearity is crucial for the statistical interpretation of quantum theory.}

Renormalization depends very much on the approach and problem at hand. It is not our aim (and beyond the scope of this manuscript) to discuss the various frameworks in detail here; instead, we shall focus purely on general properties of renormalization that are shared by the standard approaches and which are relevant to our arguments. Renormalization always requires a notion of {\it scale}. The physical nature of such a renormalization scale can differ substantially from one problem to another. Conceptually, one may broadly distinguish two manners in which a notion of scale arises.

\begin{itemize}
\item[(1)] In the first case, scale is identified in terms of the spectrum of an observable of the theory. More precisely, in this case, renormalization simply amounts to a reorganization of quantum physical degrees of freedom according to the possible values of this specific physical observable, which provides a notion of scale. Hence, it involves a decomposition of the full Hilbert space according to the (possibly improper) eigensubspaces associated to this scale observable. For example, in quantum field theory on a background space-time or in quantum many-body physics, this scale observable may be the momentum of the field modes (which is associated to a one-body operator), in which case the short-scale degrees of freedom would be the higher momentum modes, see e.g.~\cite{salmhofer, zinn-justin, rivasseau}. Note that such a definition of scale may also be used in a quantum gravity context, for instance in group field theory \cite{Oriti}, which precisely reinterprets the question of the emergence of classical space--time from loop quantum gravity as a many-body problem \cite{Oriti2}.

\item[(2)] The second case arises only in discrete systems (but we emphasize that this is not the only way in which scale can arise in discrete systems). Instead of identifying scale in terms of the spectrum of an observable, it is sometimes more appropriate in discrete systems to parametrize scale roughly with the `complexity' of the lattice or discretization under consideration. For example, in entanglement renormalization \cite{Vidal}, {including} the multi-scale entanglement renormalization ansatz (MERA), a real-space renormalization scheme developed for condensed-matter systems, scale is related to that of the background space, but in practice determined using the entanglement structure across the lattice. Short-scale degrees of freedom are defined as those which have non-negligible short- but negligible long-range (relative to real space) entanglement and thus effectively decouple from the macroscopic dynamics. Coarse-graining amounts to removing short-range entanglement and shrinks the spatial lattice, effectively reducing its `complexity'. Furthermore, in discrete quantum gravity models -- where there is no background space-time and thereby external reference scale -- the renormalization scale can be more directly identified with the complexity of the underlying discretization \cite{Bianca, BiancaSebbl,Benny}.
\end{itemize}

In both cases, coarse-graining and renormalization amount to a reduction of the `size' of the Hilbert spaces. More precisely, the irrelevant information stored in physically inaccessible scales is coarse-grained through a linear projection which restricts the state space in case (1) to lower-scale eigensubspaces of the scale observable, and in case (2) to a coarser lattice/discretization.\footnote{In fact, there exists an alternative philosophy for coarse-graining quantum systems: instead of projecting out ``irrelevant'' short-scale degrees of freedom, upon separating the relevant and irrelevant degrees of freedom, one can also work at the level of density matrices and trace out the irrelevant degrees of freedom (see, e.g.\ \cite{densityren}). However, tracing out degrees of freedom preserves linearity and thus also such an approach is compatible with our arguments.} This often involves setting the short-scale degrees of freedom into their vacuum state and `dropping' them from consideration. Neglecting (for our purposes) irrelevant details, in either case, coarse-graining schematically defines a {\it linear} map
\ba
\mathbb{P}&:&\mathcal{H}_{\rm fine}\rightarrow\mathcal{H}_{\rm coarse}\nn
\ea
from a Hilbert space representing the finer to a Hilbert space carrying the coarser degrees of freedom; those states in $\mathcal{H}_{\rm fine}$ which are operationally indistinguishable at the scale defined by $\mathcal{H}_{\rm coarse}$ define an equivalence class of states, all of which are projected to the same coarser state on $\mathcal{H}_{\rm coarse}$. If the Hilbert spaces are finite-dimensional, then $\dim\mathcal{H}_{\rm coarse}<\dim\mathcal{H}_{\rm fine}$. On the other hand, in the infinite-dimensional case, the projection will typically not reduce the Hilbert space dimension but, e.g., the dimension of the underlying configuration space over which the Hilbert space is defined. In most cases, $\mathbb{P}$ will be a {\it proper} projector which means that $\mathcal{H}_{\rm coarse}\subset\mathcal{H}_{\rm fine}$ is a proper subspace and $\mathbb{P}^2=\mathbb{P}$ is well-defined (e.g., in the finite-dimensional case or if $\mathbb{P}$ projects onto an entire interval of the continuous spectrum of a scale observable). However, for our discussion it is relevant to note that in some coarse-graining schemes for discrete systems -- in particular, for discrete quantum gravity models -- the coarse-graining steps are defined by the dynamics of the system and are tantamount to imposing constraints \cite{Bianca,icke1,icke2,icke3,BiancaDyn,BiancaSebbl,DG,BDG}. If the spectra of these constraints are continuous (around zero), $\mathbb{P}$ will define an {\it improper} projection which means that the space of solutions to the constraints (i.e., the image of $\mathbb{P}$ to be completed to $\mathcal{H}_{\rm coarse}$) is {\it not} a proper subspace of $\mathcal{H}_{\rm fine}$ \cite{raq1,raq2,thiemann,icke1,icke2,icke3}. In this case, $\mathbb{P}^2$ is not well-defined (one obtains a formal divergence ``$\mathbb{P}^2=\infty\cdot\mathbb{P}$'') and the image states will not be normalizable in the inner product on $\mathcal{H}_{\rm fine}$. But this does not affect our argument. Indeed, in either case, the linear vector space structure of $\mathcal{H}_{\rm coarse}$ is directly inherited from that of $\mathcal{H}_{\rm fine}$. 

Even stronger, the inner product on $\mathcal{H}_{\rm coarse}$ is likewise inherited from that on $\mathcal{H}_{\rm fine}$. More precisely, 
\ba
\langle\psi_{\rm coarse}|\phi_{\rm coarse}\rangle_{\rm coarse}:=\langle \psi_{\rm fine}|\mathbb{P}\,\phi_{\rm fine}\rangle_{\rm fine}\,,\label{IP}
\ea
where $\mathbb{P}(\psi_{\rm fine})=\psi_{\rm coarse}$, formally defines an inner product on $\mathcal{H}_{\rm coarse}$ in terms of that on $\mathcal{H}_{\rm fine}$. In both the proper and improper case, (\ref{IP}) does not depend on the representatives of the equivalence classes of $\phi_{\rm fine},\psi_{\rm fine}$. Namely, suppose $\psi_{\rm fine},\tilde{\psi}_{\rm fine}$ are in the same equivalence class, i.e.\ $\mathbb{P}\,\psi_{\rm fine}=\mathbb{P}\,\tilde{\psi}_{\rm fine}$. Then, since $\mathbb{P}$ must be symmetric, 
\ba
\langle \psi_{\rm fine}|\mathbb{P}\,\phi_{\rm fine}\rangle_{\rm fine}=\langle \mathbb{P}\,\psi_{\rm fine}|\phi_{\rm fine}\rangle_{\rm fine}=\langle \mathbb{P}\,\tilde{\psi}_{\rm fine}|\phi_{\rm fine}\rangle_{\rm fine}=\langle \tilde{\psi}_{\rm fine}|\mathbb{P}\,\phi_{\rm fine}\rangle_{\rm fine}.\nn
\ea

In the case that $\mathcal{H}_{\rm coarse}$ is a proper subspace of $\mathcal{H}_{\rm fine}$, this inner product coincides with the original one on $\mathcal{H}_{\rm fine}$. But also in the case that $\mathbb{P}$ is improper -- and thus $\mathcal{H}_{\rm coarse}$ fails to be a subspace of $\mathcal{H}_{\rm fine}$ -- this inner product can usually be made well-defined -- if one pays attention to various technical subtleties which are not relevant for the present discussion \cite{raq1,raq2,thiemann,icke1,icke2}.\footnote{In fact, generally, some subtle choices have to be made which must later be checked for full consistency with the physics to be described. Nevertheless, modulo such subtleties, the general Hilbert space structure of $\mathcal{H}_{\rm coarse}$ is inherited from that of $\mathcal{H}_{\rm fine}$.} For example, in the case that $\mathcal{H}_{\rm fine}$ coincides with the space of square-integrable functions over some configuration manifold, an improper $\mathbb{P}$ will typically map elements of $\mathcal{H}_{\rm fine}$ to distributions over the same configuration manifold which, together with a square integrable function, integrate to a finite value for (\ref{IP}). We emphasize, however, that the resulting inner product on $\mathcal{H}_{\rm coarse}$ in the improper case is distinct from that of $\mathcal{H}_{\rm fine}$ (after all, the image states of $\mathbb{P}$ are not normalizable in the inner product on $\mathcal{H}_{\rm fine}$).
 
Hence, in either case, Cauchy-completing the image of $\mathbb{P}$ and equipping it with the inner product (\ref{IP}) yields $\mathcal{H}_{\rm coarse}$ whose Hilbert space structure is inherited from that of the finer  $\mathcal{H}_{\rm fine}$. In this sense, the probabilistic structure of the theory survives renormalization.

Recall that we have related the inner product on the spin Hilbert space (or rather its Hermitian matrices) to the Euclidean structure of space-time in Thought experiment 2. Thus, the arguments above are crucial to admit the following interpretation: \emph{It is consistent to regard the Euclidean structure of space as arising from the Hilbert space inner product of low-energy quantum mechanics, which itself is derived from the inner product in the fundamental quantum theory of gravity.} In summary, this supports our conjecture that the local Euclidean (not only linear) spatial structure is ultimately grounded in the Euclidean structure of the states and observables of the fundamental quantum theory.

Similarly as in Section~\ref{SecStatisticalModels} and Section~\ref{SecSpin}, we can instead focus on the set of observables and its linear and Euclidean structure, without direct reference to Hilbert space. In the process of renormalization outlined above,
finer observables are linearly mapped to coarser ones. This is achieved by a linear embedding map 
\ba
\mathbb{E}:\mathcal{H}_{\rm coarse}\rightarrow \mathcal{H}_{\rm fine}\nn
\ea
which sends the coarser state back to a representative from its associated equivalence class in $\mathcal{H}_{\rm fine}$ such that $\mathbb{P}\cdot\mathbb{E}=\rm{id}_{\rm coarse}$. Depending on the system and method under consideration, there exist various ways of achieving this. For instance, this might involve putting the finer ``irrelevant'' degrees of freedom into a vacuum or other special state \cite{Vidal, Bianca, BiancaSebbl, icke1,icke2,icke3,DG,BDG}. In either case, coarse-graining the observables then proceeds schematically according to 
\ba
O_{\rm fine} \mapsto O_{\rm coarse}:=\mathbb{P}\,O_{\rm fine}\,\mathbb{E}\nn
\ea
which is {manifestly} linear. The Hilbert-Schmidt inner product on the coarse observables is inherited from the Hilbert-Schmidt inner product on the fine observables. This is the inner product that we have identified as directly operationally relevant in Section~\ref{SecStatisticalModels}.

For concreteness, let us now briefly summarize how renormalization is implemented in standard QFT; we will subsequently discuss some tentative extensions to quantum gravity which have already been proposed in the literature and are being actively investigated.

Quantum field theories on a fixed space-time background are organized according to standard energy scales, which form the spectrum of the kinetic energy operator\footnote{Note that this is a kinetic observable, hence it acquires a well-defined physical meaning only asymptotically. A full rigorous construction of interacting field theories being still missing, the coarse-graining procedure outlined in the previous paragraph operates within the Hilbert space of the free theory.}. A cut-off is moreover present to account for the fact that the theory is only applicable in a certain range of scales and is in particular expected to break down at high energy\footnote{For instance, QED does formally break down at extremely high energy -- well beyond any relevant scale of particle physics --, while QCD is asymptotically free and therefore well-behaved at arbitrarily high energy. Irrespectively of these mathematical facts, new physics is of course susceptible to show up any time uncharted territories are probed, and therefore the standard model -- and any physical theory for that matters -- is not to be trusted at arbitrarily high energy scales.}. Renormalization then amounts to an adjustment of the cut-off scale to the scale of the dynamical processes one is interested in, keeping the low energy physics fixed.  

In an effective field theory approach to quantum gravity -- such as Asymptotic Safety \cite{Reuter:2012id} --, the smooth structure of space-time is built-in, therefore the local vectorial structure of space-time is essentially assumed from the outset. We can nonetheless wonder how observables such as the angular momentum of some space-time region are realized in this approach. The dynamical treatment of the space-time metric relies on a split into a background field and fluctuations, and a notion of scale is introduced through the spectrum of the propagator of the fluctuations over the background.
The key challenge is to show that such a theory may be consistently renormalized\footnote{As proven in the seventies and eighties quantum gravity is perturbatively non-renormalizable \cite{GoroffSagnotti, ThooftVeltman}, the Asymptotic Safety programme therefore relies on and tests the existence of a non-perturbative fixed point.}, and in a way which preserves diffeomorphism invariance and background independence\footnote{This requires in particular that the so-called \textit{split symmetry} {\cite{ReuterBecker}},  which transforms both the background metric and the fluctuations, be {restored after the full renormalization group trajectory has been integrated out}.}. What is important for us is that, again, the linearity of the (asymptotic) observables is preserved by the renormalization procedure. If this programme can be brought to completion, one could therefore investigate whether or not there is a sense in which the angular momentum of, e.g., a black hole emerges from the basic linear structure of the microscopic quantum observables. 

In more radical quantum gravity approaches such as Loop Quantum Gravity \cite{thiemann}, Spin Foams \cite{alejandro} or Dynamical Triangulations \cite{cdt}, which aim at being fully background-independent from the outset, the background geometry of space-time itself has to be an outcome of the coarse-graining procedure. Even if some aspects of the local topology and geometry of space-time (as, say, a notion of Lorentz covariance in Spin Foams \cite{Rovelli}) are hard-wired at the microscopic level, it remains to be understood whether and how they give rise to the local structure of general relativity. This requires new background-independent renormalization techniques -- with a suitable background-independent notion of `scale'. 

Two such candidate formalisms are, for instance, being developed in the context of Spin Foams, which are based on two alternative interpretations of spin foam amplitudes: on the one hand, as generalized lattice regularizations of formal smooth space-time amplitudes (see \cite{Bianca, BiancaSebbl,Benny} and references therein); and, on the other hand, as quantum space-time histories to be summed over (in agreement with the superposition principle) by a non-local quantum field theory called group field theory (see \cite{Oriti, Rivasseau, Sylvain,SylvainPhD} and references therein). Interestingly, this leads again to the same dichotomy we already found in standard renormalization theory: while the scale underlying the lattice interpretation is determined by the complexity of the space-time foam, the group field theory scales are instead determined by the spectrum of a suitably chosen (one-body) observable. In both pictures, classical space-time geometry is hoped to emerge from the fundamental theory in a ``geometrogenesis'', i.e.\ a phase transition from a discrete to a condensed (smooth) geometric phase \cite{geo1,geo2,geo3}. The key point for us is that these works are producing generalizations of lattice and quantum field theory renormalization techniques which, even though they rely on different qualitative set-ups, always preserve the basic linearity of quantum gravity observables\footnote{This statement involves many subtleties and qualifiers, in particular as regards the notions of observable and scale in quantum gravity, which go beyond the scope of this article.}. 

Similarly, emergence of geometries also plays a key role in the holographic context of the AdS/CFT correspondence. While the conceptual and technical framework of this approach to quantum gravity is quite different from the previously mentioned ones, bulk geometric properties can be interpreted as also arising via a renormalization, in this case from coarse-graining data of the quantum field theory on the boundary. The radial direction corresponds to the `scale' and renormalization proceeds from the ``UV'' on the boundary to the ``IR'' in the bulk \cite{skenderis}. This holographic emergence of geometries (at least in lower dimensions) can be studied using the rich toolbox of tensor networks and (both the discrete and continuous) MERA \cite{swingle1,swingle2,nozaki}. Again, this renormalization scheme preserves linearity.

From these examples, we see that there is a sense in which the linear structure of quantum theory is preserved by any renormalization procedure, and also in quantum gravity. This is a feature that the various approaches to quantum gravity have in common -- despite the otherwise drastic conceptual and technical differences among them. Thus, if there is any grain of truth to our line of arguments, then the linearity that we see in ``every-day'' physics is a direct consequence of the probabilistic linearity of the microscopic quantum theory of nature. Moreover, in view of Thought experiment 2 in Section~\ref{SecSpin}, the spatial Euclidean metric itself can be encoded in the inner product on the space of low-energy quantum field observables, and, as suggested by our above renormalization arguments, might therefore be traced back to the fundamental inner product of the statistical model of the microscopic quantum gravity degrees of freedom. Importantly, the formulation of this conjecture makes no reference to any particular quantum theory of gravity; our arguments are rather {\it universal}, i.e.\ theory-independent, and could therefore be tested and challenged within any of them.

\section{Conclusions and outlook}
\label{SecConclusions}
Starting from the simple observation that linear spaces permeate all of our descriptions of the physical world, in particular quantum theory and general relativity, we speculate that they might have a unique origin in the probabilistic linearity of an underlying (more) fundamental theory of nature, presumably quantum gravity. We support this conjecture with several arguments from a blend of recent developments and ideas in the foundations of quantum (information) theory and quantum gravity. 

More precisely, we argued first that the fundamental linearity of quantum theory is actually its probabilistic rather than state vector linearity (``superposition principle''), the latter being implied by the former (together with the overall structure of the state space). This argument is substantiated by a recent wave of reconstructions of quantum theory \cite{Fuchs,Hardy2001,DakicBrukner,MasanesMueller,Chiribella,Hardy2011,Masanes2013,Barnum2014,Hoehn2014,Hoehn2015,Wilce2016} in which the convex set of density matrices appears from operational probabilistic structures as a primary entity, while Hilbert space structures are secondary. In fact, without loss of generality, one can directly describe quantum theory in the density matrix picture in which case the probabilistic (convex) linearity is crucial, while the state vector linearity {is secondary}. Second, not only the state vector linearity, but also the inner product on Hilbert space can be derived from probabilistic arguments. Third, we related the spatial Euclidean metric with the {Hilbert-Schmidt} inner product of the low energy quantum (field) observables and reasoned that the former is encoded in the latter. This provides a connection between linearity of local spatial geometry and probabilistic linearity. Finally, we argued that the linear Hilbert space structure of the low energy quantum degrees of freedom is inherited, through renormalization, from the Hilbert space structure of whatever the fundamental quantum theory of nature might be. Indeed, renormalization and coarse-graining are anticipated to provide the bridge from the Planck scale physics of quantum gravity to the known physics of quantum fields in classical space-times. Taken together, these arguments suggest that the local linearity of space-time structure is, in fact, a consequence of the probabilistic linearity of a (more) fundamental theory of nature.

Such a reasoning can possibly be pushed even further: using a quantum communication thought experiment, it is possible to derive the orthochronous Lorentz group $\rm{SO}^+(3,1)$ as the set of dictionaries among different observers' accounts of physics from operational consistency conditions~\cite{Hoehn}. This can be viewed as unraveling an operational quantum origin of the Lorentz group which usually is taken as a product from classical physics. Crucially, the derivation of this result does not presuppose any specific space-time structure; instead, the probabilistic linearity of the quantum formalism plays a fundamental role. It remains to be checked in more detail that the action of the Lorentz group found in~\cite{Hoehn} does coincide with the usual Lorentz transformations of space-time observers. (This is not obvious as the derivation does not adhere to any spacetime structure, but only the properties of quantum systems.) Plausibility arguments in favour of this interpretation have been proposed (such as a realization of these transformations in the context of a relativistic Stern-Gerlach measurement device) and need to be strengthened, which is the subject of ongoing efforts. If confirmed, this will suggest that not only the local linear structure of space-time, but possibly also the local Lorentz group might be regarded as a consequence, through renormalization (and basic operational arguments), of the probabilistic linearity of a (more) fundamental theory of nature. We emphasize that the previous four sections of \emph{this} paper did \emph{not} aim at saying anything about the symmetry group of space-time directly, only about its spatial linear and Euclidean structure.

The virtue of the line of arguments in this manuscript is its {\it universality}, i.e.\ independence of the precise approach to a quantum theory of gravity; it relies on universal properties of renormalization and not on the precise degrees of freedom that are being coarse-grained. In fact, such universal arguments could be a precious guiding thread and incentive to develop alternative `top-down' approaches to the conundrum of quantum gravity (where `top/down' refers to large/small length scales), aspiring to elucidate how (some of) the known classical space-time structures may emerge {\it universally} from whatever the precise microphysics are. Such a `top-down' approach could complement existing `bottom-up' approaches that, instead, start with a choice of microscopic degrees of freedom and face the daunting task of reproducing the known physics at large scales.

Indeed, there is good evidence that it actually might be worthwhile to follow such a `top-down' perspective on quantum gravity, in addition to more standard `bottom-up' approaches. For example, another seemingly universal property of quantum gravity, receiving much attention in the literature recently, are area laws for the entanglement entropy of vacuum-like (Hadamard) states of a quantum field across arbitrary semiclassical space-time regions. It has indeed been conjectured \cite{bianchi} that such an area law constitutes a characteristic signature of the emergence of a semiclassical geometry in a quantum theory of gravity. The arguments supporting this conjecture, in analogy to our own, require a suitable coarse-graining of the UV physics. The reason these conjectured universal properties are interesting is that they likely hold the key to the derivation of the correct semiclassical gravitational dynamics. Namely, the semiclassical Einstein equations can be derived from such area laws and entanglement entropy equilibrium conditions \cite{ted} (for an analogous derivation of the linearized Einstein equations within the AdS/CFT correspondence, see \cite{raamsdonk}).

Despite being ignorant of the precise microphysics, universal arguments of the kind discussed here contribute to closing the gap between whatever the putative quantum theory of gravity may be and the known large-scale physics. Specifically, constraining the fundamental theory to produce emergent metric geometries with area laws seems to be an `easier' task than directly asking it to generate the semiclassical Einstein equations, especially given the evidence that area laws are quite generic for ground states of complex quantum systems~\cite{eisert}.

In this spirit, we hope that the proposal made in this article may help to shed further light on the emergence of such geometric structures -- such as inner product, local linearity and Lorentzian structure -- from properties of an underlying microscopic quantum theory of nature. In particular, we note that the derivation of the semiclassical Einstein equations from area laws already presupposes an underlying local linear (Lorentzian) space-time structure as a semiclassical limit of some quantum theory of gravity. This proposal may thereby provide additional impetus for justifying this crucial assumption.

If there is a grain of truth to the hypothesis that we have put forward in this paper, then this has a rather amusing consequence for our understanding of the architecture of physics. In the Philosophy of Physics, especially in some approaches to the interpretation of quantum mechanics, it is often stated that (quantum or classical) probabilities should be viewed as ``degrees of belief'', or knowledge, of agents, rather than actual, objective properties of the world. This is most famously expressed in de Finetti's quotation that ``Probabilities do not exist''~\cite{Finetti}. However, if some of the structure of space-time derives from the structure of probability theory, as we have argued above, then this holds a quite different message: it tells us that some aspects of physics that we have so far considered as hard ``elements of reality'' (such as, possibly, tangent space) may in fact be no more fundamental than probability itself.

\section*{Acknowledgments}
We are grateful to Bianca Dittrich and Wolfgang Wieland for  discussions and helpful feedback on an earlier draft. This research was supported in part by Perimeter Institute for Theoretical Physics. Research at Perimeter Institute is supported by the Government of Canada through the Department of Innovation, Science and Economic Development Canada and by the Province of Ontario through the Ministry of Research, Innovation and Science. MPM acknowledges funding from the Canada Research Chairs program. The project leading to this publication has also received funding from the European Union's Horizon 2020 research and innovation programme under the Marie Sklodowska-Curie grant agreement No 657661 (awarded to PAH).

\end{document}